# Gate modulation and interface engineering on Coulomb blockade in open superconducting islands

Huading Song[1,*,†], Dong Pan[2,*], Runan Shang[1,3,‡], Zhaoyu Wang[4], Ke He[1,3,4,5], Jianhua Zhao[2] and Hao Zhang[1,4,5,§]

[1]*Beijing Academy of Quantum Information Sciences, Beijing 100193, China*
[2]*State Key Laboratory of Superlattices and Microstructures, Institute of Semiconductors, Chinese Academy of Sciences, P.O. Box 912, Beijing 100083, China*
[3]*Hefei National Laboratory, Hefei 230088, China*
[4]*State Key Laboratory of Low Dimensional Quantum Physics, Department of Physics, Tsinghua University, Beijing 100084, China*
[5]*Frontier Science Center for Quantum Information, Beijing 100084, China*

**Abstract:** Mesoscopic Coulomb blockade (MCB) is recognized as a phase-coherent variant of the conventional Coulomb blockade that arises in systems with open contacts. In open quantum dots, MCB is enhanced by a decrease in background conductance. This occurs because the reduction in coupling strength between the quantum dot and the outer reservoir renders the system more closed, thereby facilitating the emergence of conventional Coulomb blockade. In this work, we demonstrate that the MCB in open superconducting islands exhibits an different correlation with coupling strength compared to open quantum dots. Specifically, a decrease in background conductance may result in a weakening of the MCB. This observation indicates that the MCB in superconducting islands originates from the presence of superconducting-normal interfaces.

## 1. Introduction

The Coulomb blockade effect is commonly observed in quantum dot systems that are weakly coupled to external reservoirs through tunneling coupling, demonstrating localization of a single-electron charge [1, 2]. Conversely, in open systems where the quantum dot contacts possess a nearly unit transmission probability, charge can move freely without restrictions, leading to almost vanishing charging energy. Studies have shown that even in open systems, quantum interference can still cause periodic oscillations in conductance with charge quantization, known as mesoscopic Coulomb blockade (MCB) [3-5]. However, there has been an absence of experimental evidence to introduce MCB into the superconducting analogue of quantum dots, namely, the open superconducting island system [6-7]. Recently, our work has successfully filled this missing part. We report the first observation of MCB in a fully open superconducting island based on ultra-thin InAs-Al nanowires [8]. These semiconductor-superconducting nanowires have garnered significant attention due to their potential to realize Majorana zero modes [9,10] and topological qubits [11-14]. The 2$e$-period oscillation corresponding to cooper pairs is observed, and the effects of magnetic field and temperature have been meticulously investigated. However, uncertainties persist regarding the correlation between MCB and coupling strength, and whether pure InAs nanowires can induce such phenomenon.

---

[*] These authors contributed equally to this work.
 Corresponding authors: [†]E-mail: songhd@baqis.ac.cn; [‡]E-mail: shangrn@baqis.ac.cn; [§]E-mail: hzquantum@mail.tsinghua.edu.cn

In this work, we systematically investigate the influence of coupling strength and device interfaces on MCB. Our findings reveal that while gate control allows for the adjustment of coupling strength, reducing the coupling does not enhance the MCB as might be expected. On the contrary, it may substantially suppresses the oscillation amplitude and can even lead to the complete disappearance of MCB - a behavior strikingly different from that observed in open quantum dots [3, 4]. Through interface engineering of the devices, we provide compelling evidence that the MCB in open superconducting islands arises from the presence of superconductor-normal interfaces. This work enhances our understanding of MCB in open superconducting islands and offers valuable insights for future research aimed at integrating MCB with signatures of topological superconducting transport [15-17]. Our results will also provide guidance for the precise calibration of the parameter space to identify the existence of

Majorana zero modes [18-20].

## 2. Methods

The ultra-thin InAs-Al nanowires (diameter ~ 40nm) were grown by in-situ molecular-beam epitaxy [21]. Device A [Fig. 1(a)] was fabricated with a four-terminal setup, containing four Ti/Au electrical contacts : N1, N2, S1, and S2. The device has two side gates, TG and SG, and one global back gate, BG. The segment between N2 and S1 is composed of a pure InAs tunneling barrier region connected in series with an InAs-Al open superconducting island region. For scanning electron microscopy image and superconducting gap characterization of device A, see Fig. S1. The coupling strength on the left can be tuned by side gate TG. Due to the huge electron concentration in the Ti/Au electrode, which can essentially pin the Fermi level at the contact S1, the coupling strength at the right can be considered as a fixed strong coupling. Gate SG was used to tune the chemical potential of the device. The substrate is p-doped Si (BG) covered by 300-nm-thick $SiO_2$. The devices were measured with temperature ~20 mK in dilution fridge. A bias DC voltage $V$ together with a lock-in excitation was applied to contact N1. Four-terminal voltage $dV$ between N2 and S2 was measured by a lock-in amplifier to exclude the contact resistance. The current $dI$ was drained from S2 and detected by another lock-in amplifier. The differential conductance $dI/dV$ was calculated from the results above. For details of the measurement setup, see Refs.[13, 14].

## 3. Results and analysis

Figures 1(b) shows that zero-bias conductance displays oscillations with stable period on $V_{SG}$ ( ~ 10mV). Simultaneously, the bias spectroscopy presents Coulomb diamond patterns. These phenomena clearly testify to the emergence of MCB. Figure 1(c) exhibits evolution of bias scans within an identical MCB period, the conductance near zero bias exhibits a distinct periodic evolution pattern, showing a clear conductance suppression at the center of the Coulomb diamond. The conductance is not suppressed to almost zero like conventional Coulomb blockade [1,2].

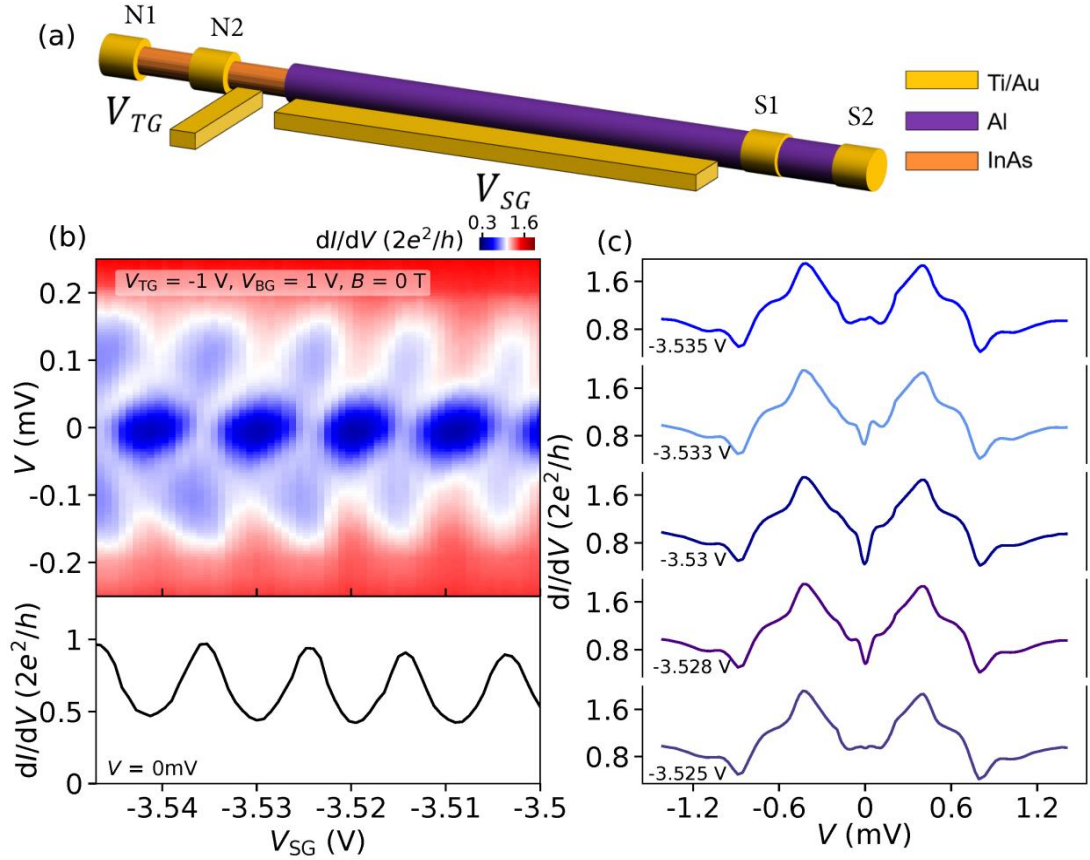

FIG. 1. (a) Diagram of the four-terminal InAs-Al nanowire device. (b) dI/dV as a function of $V_{SG}$ and $V$, demonstrating MCB oscillations and Coulomb diamonds. The lower panels show line cuts at zero bias. (c) Horizontal line cuts in an identical oscillation period.

Due to the crosstalk between the TG, SG and BG, the coupling strength is modulated by all three gates [13]. In Fig. S2, we show tuning TG alone does not have a significant impact on MCB, as background conductance is stable. Consequently, we propose that the smoothed background conductance more accurately reflects the coupling strength. Figures 2(a,b) illustrate the evolution of background conductance and oscillation signals over a $V_{SG}$ scan. Consistent with our previous work, the oscillation period shifts from 1$e$ [Fig. 2(a)] to 2$e$-period [Fig. 2(b)].

A clear trend emerges: peaks in background conductance generally coincide with regions of stronger oscillation signals, whereas valleys are associated with a decrease or even a complete disappearance of oscillation signals. To quantitatively assess this relationship, we performed FFT analysis on the oscillation signals in the peak (red) and valley (blue) regions of background conductance. The results are displayed in Figs. 2(c,d). In the 1*e* oscillation region, the FFT result for the peak region reveals a dominant peak at 190.3 $V^{-1}$, corresponding to a period of ~ 5 mV, which closely aligns with the oscillation period depicted in Fig. 2(c). In the 2*e* oscillation region, the dominant peak occurs at 93 $V^{-1}$, corresponding to a period of ~ 10 mV. The MCB is significantly pronounced in the peak regions of background conductance, while it is notably suppressed in the valley regions. These findings indicate a positive correlation between the MCB and the coupling strength, contrasting sharply with observations in closed quantum dot systems [3,4]. In the latter case, the MCB becomes more pronounced as both the coupling strength and background conductance decrease, approaching the conventional Coulomb blockade characteristic of closed quantum dots.

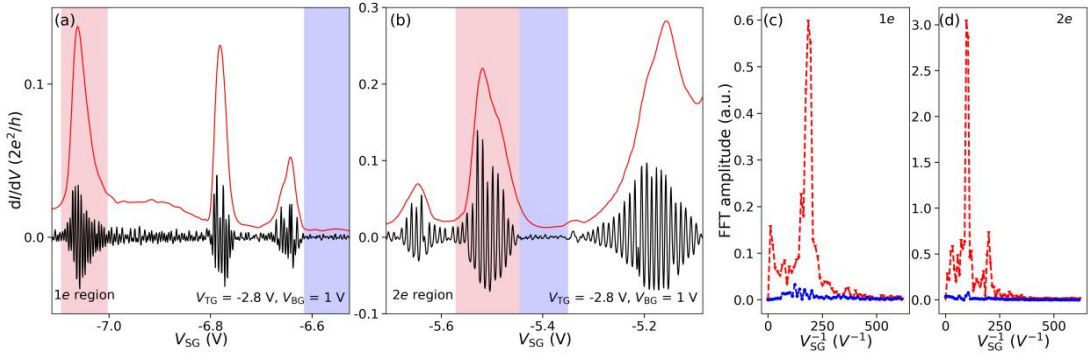

FIG. 2. (a,b) Oscillation signal and smoothed background conductance as a function of $V_{SG}$ in 1*e* and 2*e* oscillation regimes. (c,d) FFT analysis results of the color shadowed peak and valley regions.

In Figure 3a, we present the smoothed background conductance and oscillation signals from a wide-range scan of SG, which includes the 1*e* oscillation region and a transition region between 1*e*/2*e*. Overall, the oscillation signals in the peak regions are enhanced compared to the adjacent valley regions. However, in some peak regions, this enhancement is not very pronounced. Additionally, a higher value of background conductance does not necessarily imply a greater degree of enhancement in the amplitude of the oscillation signal. We propose that the occupation of multi-subband in InAs-Al region, the emergence of Andreev bound states, and the quasiparticle poisoning may have impact on this phenomenon [16, 22-26]. Figures (3c,d) show two zoom-in views of the 1$e$ oscillation and transition region from (a), clearly revealing that the amplitudes of the neighboring oscillation peaks in the 1$e$ region are quite similar. Whereas in the transition region, there is an obvious difference in the amplitude between the 1$e$ and 2$e$ states. The oscillation characteristics in transition region are highly consistent with the quasi-particle poisoning in previous nanowire superconducting island experiments [15, 16]. We propose that the degree of quasi-particle poisoning can be tuned by $V_{SG}$, leading to the modulation of even-odd asymmetry in the system's ground state energy (Fig. S3). The emergence of Andreev bound states can also induce the 1$e$-2$e$ transition [16, 26]. The FFT results are displayed in Figs. 3(d,e). The FFT peak of 1$e$-region is at 190.4 $V^{-1}$. In the transition region, two dominant peaks occurs at 91.3 and 183.3 $V^{-1}$, corresponding to the observation of both periods. The FFT amplitude in peak regions is still obviously enlarged compared to that of valley regions.

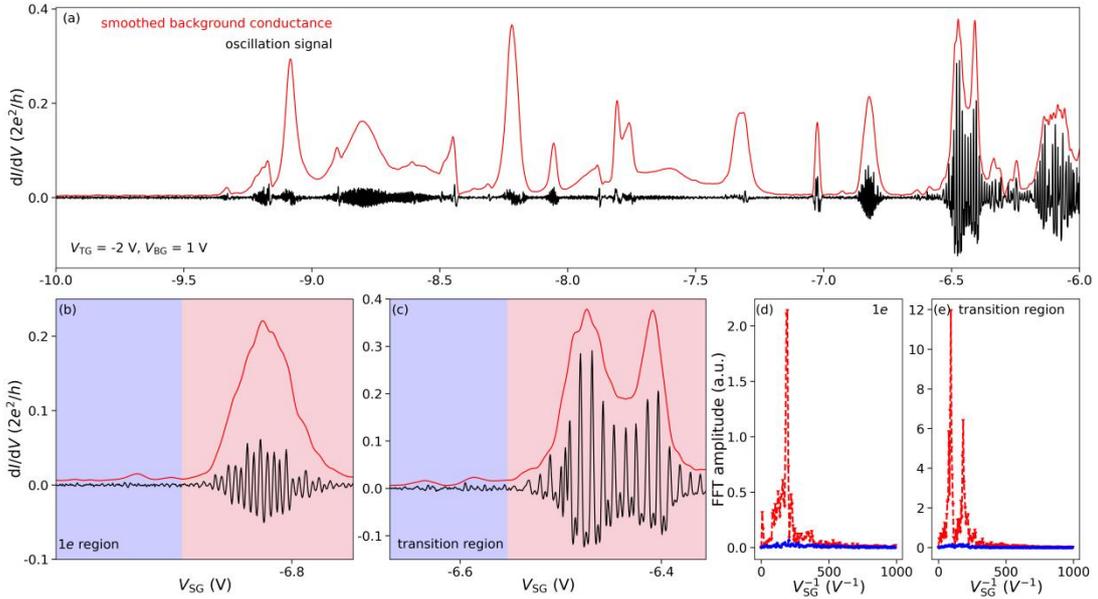

FIG. 3. (a,b) Oscillation signal and smoothed background conductance as a function of $V_{SG}$ in a broad scan of $V_{SG}$. (c,d) Zoomed-in views of the 1$e$ oscillation and transition regions. (e,f) FFT analysis results of the color shadowed sections.

To understand the correlation between MCB and the coupling strength, we present measurements obtained from two alternative layouts. The first layout consists of an InAs-Al short structure that makes direct contact with Ti/Au electrodes, while the second layout features a pure InAs segment. Both devices are fabricated on the same nanowire. Under zero magnetic field, the former device exhibits conductance oscillations with a uniform period [Fig. 4(a)], whereas the latter fails to display regular oscillations [Fig. 4(b)]. The smoothed background conductance is shown in red curve. Figures 4(c,d) show the oscillation signals from Figure 4(a) and the corresponding FFT analysis results. The oscillation period is 260 mV, greatly enlarged due to the small size of the device. The dominated FFT peak is at 3.85 $V^{-1}$. These findings confirm that the MCB originates from the superconducting-normal interfaces. We propose a physical mechanism for this phenomenon, illustrated in Fig. 4(e), wherein Cooper pairs undergo Andreev reflection at the two superconducting-normal interfaces. It is important to note that the superconducting island is phase coherent. The quantum interference between Andreev reflection paths serves to localize the charges on the island, resulting in an Andreev version of MCB. If the coupling strength on the left side decreases, it becomes more difficult for charges to enter the InAs-Al region, thereby significantly suppressing the amplitude of the MCB.

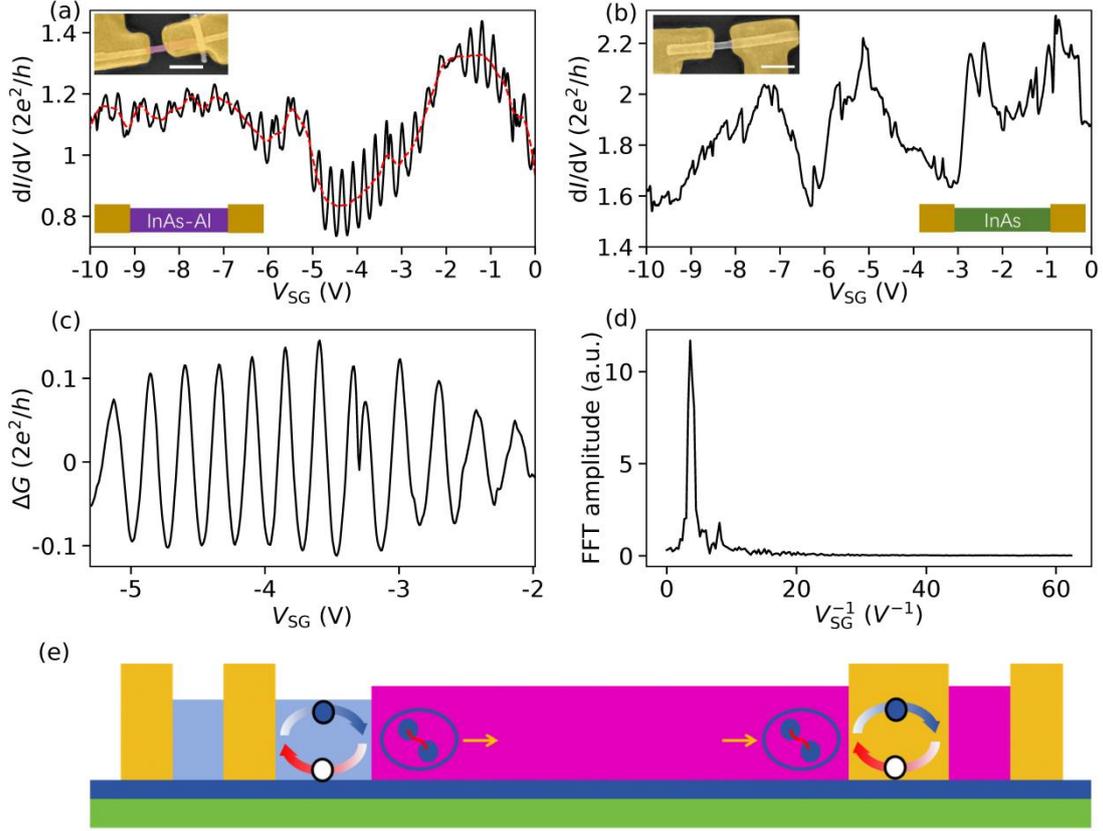

FIG. 4. (a,b) Two-terminal dI/dV as a function of back gate voltage ($V_{BG}$) in a very short InAs-Al segment and a pure InAs segment. The insets show SEM images and diagrams of the two devices. Scale bars are both 200nm. (c) A section of (a) shows periodic conductance oscillations. (d) FFT analysis shows an amplitude peak at ~ 3.85 $V^{-1}$, corresponding to stable period of 260 mV. (e) Diagram for mechanism of the MCB.

## 4. Conclusion

In conclusion, this research investigates explored the influence of coupling strength and device interfaces on MCB in open superconducting islands. Unlike open quantum dots, reducing coupling strength between open superconducting island and outer leads can suppress the MCB, and may even cause it to disappear entirely. We provide conclusive evidence that the MCB in open superconducting islands is attributed to the presence of superconducting-normal interfaces. This work enhances our understanding of MCB in open superconducting islands and may pave the way for new developments in superconducting quantum devices and related quantum information technologies.

**Acknowledgment**
This work is supported by Beijing Natural Science Foundation (Grant No.1252037), National Natural Science Foundation of China (Grants No. 12374459, No. 61974138, No. 92065206, No. 12374158, No. 12074039, No. 11974198), the Innovation Program for Quantum Science and Technology (Grant 2021ZD0302400). D. P. also acknowledges the support from Youth Innovation Promotion Association, Chinese Academy of Sciences (No. 2017156, No. Y2021043).

# Supplemental Material for "Gate modulation and interface engineering on Coulomb blockade in open superconducting islands"

Huading Song[1,*,†], Dong Pan[2,*], Runan Shang[1,3,‡], Zhaoyu Wang[4], Ke He[1,3,4,5], Jianhua Zhao[2] and Hao Zhang[1,4,5,§]

[1] *Beijing Academy of Quantum Information Sciences, Beijing 100193, China*
[2] *State Key Laboratory of Superlattices and Microstructures, Institute of Semiconductors, Chinese Academy of Sciences, P.O. Box 912, Beijing 100083, China*
[3] *Hefei National Laboratory, Hefei 230088, China*
[4] *State Key Laboratory of Low Dimensional Quantum Physics, Department of Physics, Tsinghua University, Beijing 100084, China*
[5] *Frontier Science Center for Quantum Information, Beijing 100084, China*

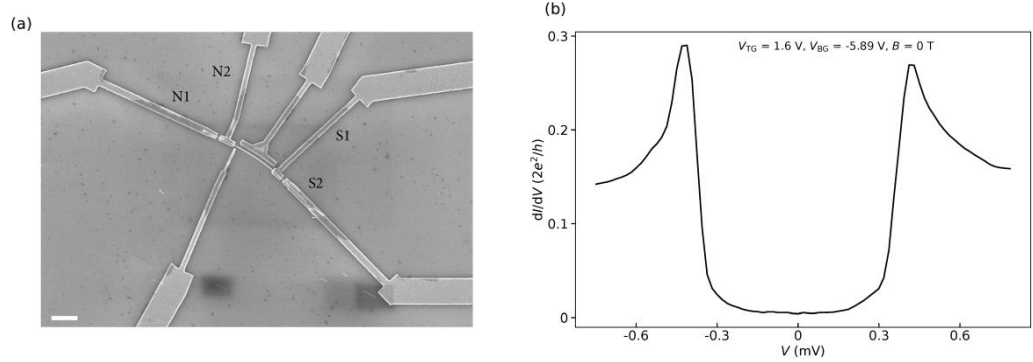

Fig. S1: (a) SEM image of device A. Scale bar: 1μm. (b) Bias scan of device A exhibits an obvious hard superconducting gap.

[*] These authors contributed equally to this work.

Corresponding authors: [†]E-mail: songhd@baqis.ac.cn; [‡]E-mail: shangrn@baqis.ac.cn; [§]E-mail: hzquantum@mail.tsinghua.edu.cn

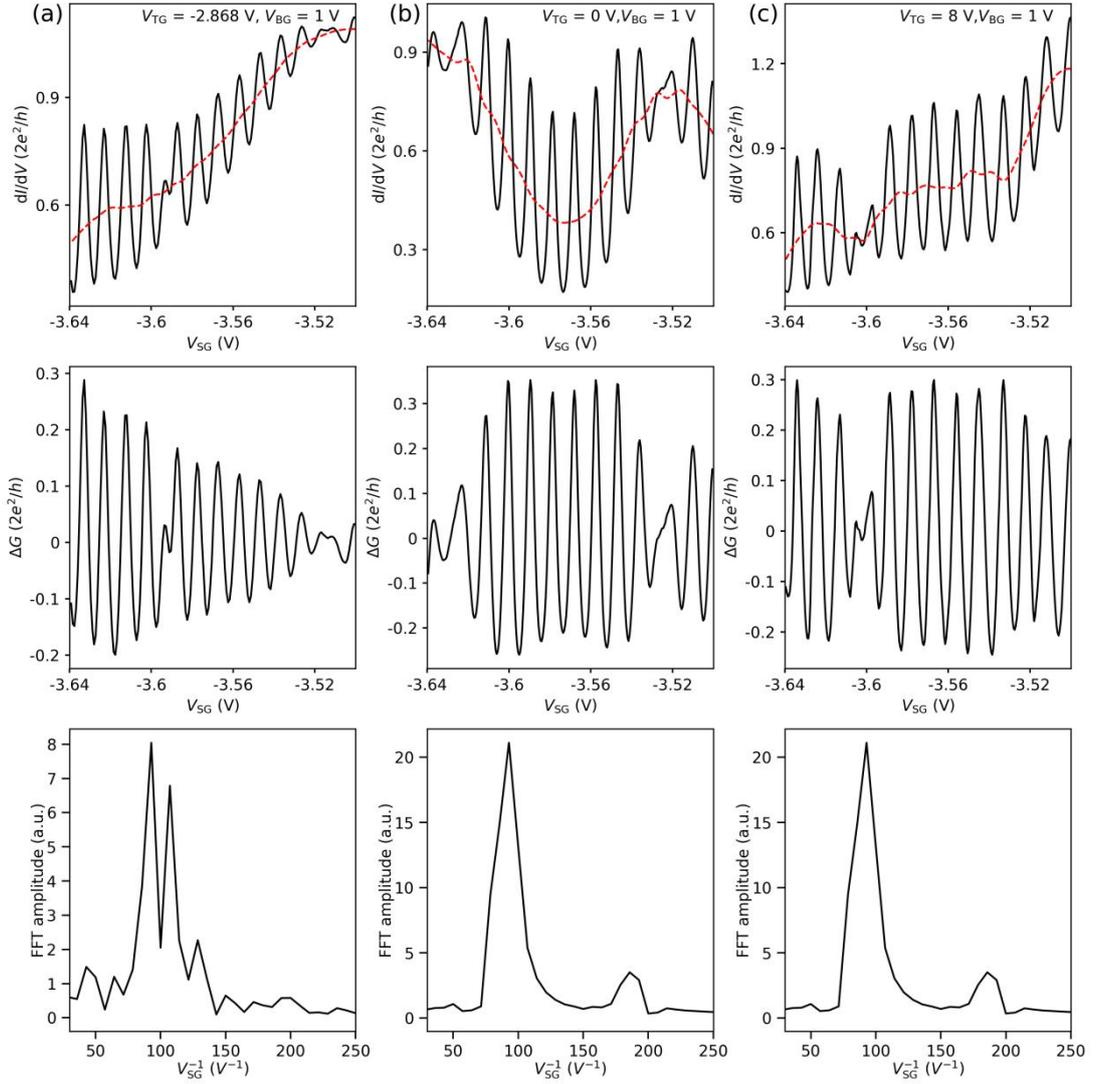

Fig. S2: Oscillation signal and corresponding FFT analysis with $V_{TG} = -2.86$ V, $0$ V, $8$ V. The scanning range of $V_{SG}$ and the values of $V_{BG}$ are kept constant. Smoothing process of the original data (top panels) was performed using Savgol filter. The oscillation signals (middle panels) were acquired after subtracting the smoothed background conductance from original data. The FFT results (bottom panels) all shows the dominant oscillation period is ~ 10 mV, corresponding to a FFT peak at ~ $96 V^{-1}$. Additionally, there is no significant change in the amplitude of the oscillation signals $\Delta G$ across the three sets of $V_{TG}$ values.

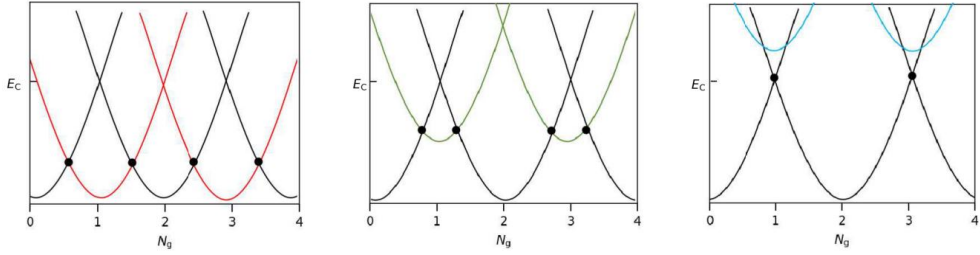

Fig. S3: Charge-state energies of a superconducting nanowire island as a function of gate induced charge $N$, with red, green and blue curves show the energy evolution of odd-parity states with different degrees of quasi-particle poisoning. The black circles mark the crossing points between odd-even parities. The black curves show energy of the even-parity states, the crossing points of adjacent curves correspond to the charging energy. The conductance peaks in the left panel occur with the same peak spacing, indicating the even-odd parity asymmetry is removed by quasi-particle poisoning. In the transition region (middle panel), the odd-parity states requires a excitation energy smaller than the charging energy, exhibiting non-equal odd ($1e$) and even ($2e$) peak spacing with larger amplitude on $2e$ oscillation peaks. In the right panel, the even-odd parity asymmetry is maintained and the odd-parity states requires a excitation energy larger than the charging energy, exhibiting only $2e$ peaks.